# Title: Interactivity: the missing link between virtual reality technology and drug discovery pipelines

Authors: Rebecca K. Walters (1), Ella M. Gale (1), Jonathan Barnoud (1), David R. Glowacki (2) and Adrian J. Mulholland (1)

((1) Centre for Computational Chemistry, School of Chemistry, University of Bristol, Bristol, UK, (2) CiTIUS Intelligent Technologies Research Centre, Santiago de Compostela 15705, Spain)

The potential of virtual reality (VR) to contribute to drug design and development has been recognised for many years. Hardware and software developments now mean that this potential is beginning to be realised, and VR methods are being actively used in this sphere. A recent advance is to use VR not only to visualise and interact with molecular structures, but also to interact with molecular dynamics simulations of 'on the fly' (interactive molecular dynamics in VR, IMD-VR), which is useful not only for flexible docking but also to examine binding processes and conformational changes. iMD-VR has been shown to be useful for creating complexes of ligands bound to target proteins, *e.g.*, recently applied to peptide inhibitors of the SARS-CoV-2 main protease. In this review, we use the term 'interactive VR' to refer to software where interactivity is an inherent part of the user VR experience *e.g.*, in making structural modifications or interacting with a physically rigorous molecular dynamics (MD) simulation, as opposed to simply using VR controllers to rotate and translate the molecule for enhanced visualisation. Here, we describe these methods and their application to problems relevant to drug discovery, highlighting the possibilities that they offer in this arena. We suggest that the ease of viewing and manipulating molecular structures and dynamics, and the ability to modify structures on the fly (*e.g.*, adding or deleting atoms) makes modern interactive VR a valuable tool to add to the armoury of drug development methods.

## 1. Introduction

Human perception, intuition, creativity, and expertise are central to computer-aided drug discovery and design (CADD) [1-5]. This is likely to remain so for the foreseeable future, despite recent significant and rapid advances in machine learning and artificial intelligence. The three-dimensional (3D) structure of molecules is essential to their function and recently drugs have started to 'escape from flatland' as molecules with more complex 3D structure are found higher up the drug development pathway due to the better solubility [6] and lower toxicity [7] they exhibited. The human mind has evolved to visualize and understand 3D space, and to interact with and manipulate objects in this world. In psychologist and computer scientist J. C. R. Licklider's 1960 essay 'Man-computer symbiosis' [8], he postulates that humans and computers will develop a symbiotic relationship, where the different strengths will complement each other, e.g., human spatial reasoning and computer speed and accuracy.

Many CADD applications that examine molecular models, allow modification of structures, and run simulations of 3D biomolecules use two-dimensional (2D) interfaces such as a monitor, using controls such as a mouse, meaning we cannot fully utilise our human 3D intelligence. Virtual reality (VR) provides a platform for 3D visualisation of complex biomolecular structures and, with the addition of interactivity via using our hands and fine motor control to directly manipulate objects, allows ways to interact, such as modifying structures 'on the fly' (*e.g.* chemical changes) or directing simulations towards a solution easily visible to a human. Thus, interactive VR is intuitive to use and enables humans to focus on the areas of the drug discovery process that benefit from human intuition, *e.g.*, chemical intuition, visualizing chirality, predicting how proteins and ligands fit together, the effects of a conformational change, or which alterations to the chemical structure might improve affinity or specificity. This should also speed development.

VR is the use of artificial sense stimuli and manual interaction with a computer program to 'trick' our mind into both perceiving a virtual world and feeling embodied in that world [9]. It allows for the addition of extra modalities, such as depth, touch, and sound, to human interaction with computer software. To be fully immersed in VR is to feel disconnected from real-world stimuli, and to feel fully engaged with the simulated world [10]. Although VR has been around for many years [11-16], recent improvements to hardware and software have made VR a technology that is ready for widespread scientific use. Advances include low cost, high resolution, fast refresh screens (for visualising VR *e.g.,* through lenses in a VR headset), fast graphics processing units (GPUs), high-level 3D graphics engines, and scientific VR software. Some examples of using VR in science and engineering for 3D visualisation are: virtual restoration of archaeological finds [17]; virtual exploration and cartography [18]; viewing the sea bed [19]; safety training in chemical manufacturing [20]; sport psychology [21]; telepresence for clinicians [22]; teaching anatomy [23,24]; investigating the molecular structures related to the SARS-CoV-2 virus [25] to name a few.

The promise of VR in drug discovery has been recognised for many years [26-28] and offers several potential benefits. Firstly, VR allows the researcher to visualise drug molecules and their macromolecular targets in full 3D, which allows a deeper understanding of these complex systems, and thus informs design and modification of ligands in the process of structure-based design and development [29,30]. Secondly, VR allows interaction with molecules through VR controllers. The controllers can be thought of as a kind of 'virtual pair of hands', allowing the user to grasp parts of a molecule or molecules as easily as if they were tangible, real-world objects. Thirdly, recent developments of interactive VR allow users to interact with a running molecular dynamics (MD) simulation on the atomic level [31], allowing them to manipulate the system, modifying its structure and interactions 'on the fly' (Fig. 1). Finally, some VR software allows multiple users to occupy the same virtual space [30,31] for collaboration and, *e.g.,* for teaching (collaborative VR). In other words, modern VR technology (hardware and software) is not merely an update with better graphics: the addition of molecular motion via MD simulations, the ability to directly manipulate those simulations, to work together virtually, and to create and modify molecular structure from within the program turns VR from a visualisation method to a research tool in its own right.

We use the phrase 'interactive VR' to refer to programs that allow the user to *interact with and change the molecular system*, and not those programs where the interactivity is limited to merely using the controllers to move or rotate the view. In this review, we discuss current VR hardware and some relevant VR software relevant for structure-based drug design. We give examples of how interactive VR is currently being used to visualise and manipulate biomolecules, addressing features that are useful in terms of CADD.

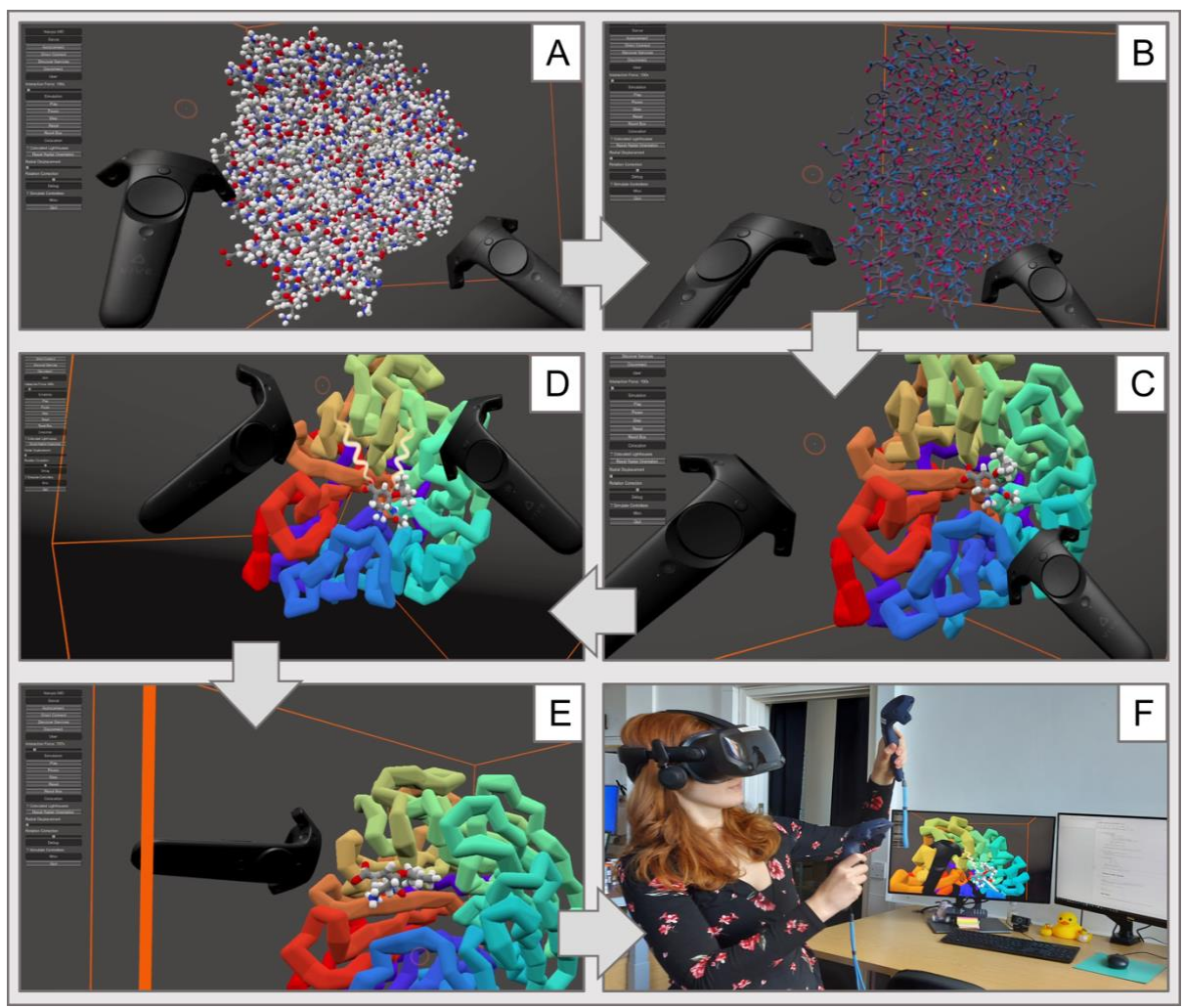

*Figure 1: **Interactivity in the VR software Narupa IMD [31,32].** A) A protein-drug system of influenza neuraminidase complexed with the neuraminidase inhibitor oseltamivir loaded into the interactive molecular dynamics in virtual reality (IMD-VR) software Narupa IMD, where all atoms are rendered by the default ball and stick renderer. The VR controllers are also rendered in VR, where the orange circle represents the point of interaction from the controller (like a cursor on a mouse). B) Rendering of all atoms changed from, ball and stick to lines and a different colour scheme. C) The protein is now rendered in the 'cartoon' renderer and coloured in rainbow, where red represents the start of the protein and violet represents the end. The drug is now rendered in the 'cpm' renderer and using atomistic colours to differentiate it from the protein. D) By grabbing single atoms from the drug molecule using each VR controller, the user can apply a force to coax the drug out of the active site. The force from the VR controllers is shown by the two light yellow sine waves between the controller and the drug molecule. E) The drug has now started to unbind from the protein following interaction from the user in VR. F) A view of the VR user in real life, where the monitor is showing what the user sees in VR.*

## 2. VR Hardware and Software

There are many different types of hardware currently available that provide a VR platform, with varying levels of immersion and interactivity. Low-cost solutions such as cardboard viewers that hold a smartphone in front of the user's eyes offer a low-cost way of experiencing VR [33-35], however, the field of view is restricted in this case, and there is limited interactivity. At the expensive end of the spectrum lies the Cave Automatic Virtual Environment (CAVE), a room-size virtual environment first developed 30 years ago that tracks the user via CAVE controllers and glasses [36] (there are also state of the art VR systems that track the user's entire body from head to feet such as Vicon Origin [37], but these are primarily used for entertainment applications so are not discussed here). A CAVE consists of multiple screens that project the simulated space onto either 4 or 6 walls. The user(s) experiences VR through stereoscopic LCD shutter glasses that display a 3D image and can also interact using CAVE compatible wireless controllers [38]. CAVE set-ups have been used for drug discovery in the past, however they are expensive, costing hundreds of thousands of dollars for a high-end set-up [39,40], and can have issues with immersion if the glasses do not cover the peripheral vision. Head-mounted displays (HMD), such as the Oculus Quest 2, HTC Vive Pro, and the Valve Index, are generally used with the VR software discussed in this review. The advantages of using HMDs are that; (i) they are now relatively cheap (between $300-$1,600 USD at the time of writing for all the necessary VR equipment, not including the computer) [41-43]; (ii) they are portable and lightweight; (iii) users experience a wider field of view than the smartphone and CAVE (ranging between 90 and 130 degrees in the horizontal dimension); (iv) they have good spatiotemporal resolution; and (v) the experience can be fully immersive (more immersive than a CAVE system, which can have issues with ventilation and claustrophobia [44]).

Typically, HMDs allow either 3- or 6- degrees of freedom (DoF). HMDs with 3-DoF allow tracking of rotational motion about the $x$, $y$, and $z$ axes (known as pitch, yaw, and roll), but do not track translational movement. Examples of 3-DoF HMDs include the Oculus Go, Samsung Gear VR, and Google Daydream. This is the simplest kind of tracking and is achieved by sensors built into the headset. HMDs that allow 6-DoF track both rotational and translational motion in the $x$, $y$, and $z$ directions. 6-DoF HMDs may require external trackers positioned within the room, while the newer HMDs, such as the Oculus Quest 2 and Oculus Rift S, are capable of 6-DoF through sensors and image processing within the headset. This kind of tracking approach seems to be the most common with the latest HMDs, indicating this is where the industry is heading. The advantage of utilizing 6-DoF in VR is that the user can exploit their full range of motion to explore the simulated environment, by walking, crouching, and bending to achieve the best view.

There are several articles that discuss and compare the specifications of the most common HMDs in detail, including resolution, refresh rate, and pixel density [45,46]. For biomolecular modelling and manipulation, a commodity HMD is often the most affordable and practical solution; current examples include the HTC Vive Pro [47], Valve Index [48], and Oculus Quest 2 [49]. The Google Daydream was recently discontinued [50]. Facebook recently renamed itself to Meta [51], after the 'metaverse' (first coined in the novel Snow Crash [52] as an online VR world accessible and manipulable by users), indicating a high likelihood of continuing production of their VR range, Oculus. Valve [48] owns the gaming platform Steam™ and is therefore likely to maintain their headset line so as to be able to continue with their core business, and HTC has been in the area for a long time having made collaborations with large corporations. All three of these headsets are reasonable choices, however we recommend the user checks whether their HMD is compatible with the interactive VR program they wish to use. Different programs mentioned in this review offer different capabilities. Some allow topological changes to the protein or ligand, some offer different renderers, some allow interaction with a physically rigorous MD simulation directly, from within the VR environment.

# 3. Tools for interactive VR: Controllers, gloves, and hands

There is a natural desire to reach out and touch simulated objects when they are presented in an immersive setting. VR controllers, such as those which can be bought as a set with HMDs, provide the user with a virtual 'pair of hands'. The user can interactively and intuitively grasp and manipulate structures as if they were physical objects in the real world. This allows a more extensive exploration of structures than manipulation on a 2D screen [53] as it is facilitated by the 6 degrees of rotational and translational freedom and the depth perception afforded by VR (although it should be noted that attempts to improve visualisation of 3D molecules have been made without the use of VR, *e.g.*, stereo viewers and 3D printing of molecules [54,55]). Compared to using a mouse to interact with the simulation, which only permits a single interaction at a time via a mouse click, some VR software allow interaction from two controllers per user simultaneously, resulting in greater control over the simulation. This can be important when exploring large molecules or studying an area of the system that may be non-trivial to navigate *e.g.*, a buried binding cavity.

However, there is still the limitation that a VR controller does not have the same range of motion as a real pair of hands, provided by the flexibility of human thumbs, fingers, and wrists. The development of VR gloves allows the incorporation of the complex motions that hands are capable of. There has been some advancement in VR gloves specifically for molecular manipulation [56].

Hand tracking is an alternative approach. Molecular Rift [57] is an example of VR software that is focused on manipulation of biomolecules in 3D space using hand tracking rather than VR controllers. Their gesture recognition software [58] allows basic structure manipulation such as rotation and translation of the molecule through a series of predetermined hand gestures. While hand tracking has the potential to be the most instinctive way for humans to manipulate molecules in VR—it removes the need for using specific hardware—a focus group of Molecular Rift users found learning the gestures to be a lengthy process as some were not immediately obvious [57].

An exciting direction is the incorporation haptic feedback (*i.e.,* the modality of touch) into the virtual world [59,60]. Specialist haptic systems have already been developed [61-67], however they can be very expensive or not easily available to the consumer. Haptic feedback in VR has been explored for use in surgical simulations [68-70], where a prospective surgeon can, for example, feel the resistance of body tissue beneath their simulated surgical tool. As this kind of physical sensation is well defined, it is relatively easy to mimic in a VR setting. Haptic feedback in terms of molecular simulation presents a different type of challenge [71-73] because there is no reference for what kind of feedback is 'realistic' or intuitively useful when manipulating these objects. On the other hand, there is the possibility of exploiting a phenomenon known as *'pseudo-haptics'* to 'feel' molecules in VR. This phenomenon is where the user experiences the *illusion* of feeling things within the simulation [73]; there is no actual physical force. A recent experiment showed that users of the interactive VR software Narupa IMD (see below) could distinguish molecular properties using this pseudo-haptic feedback [74]. The study involved users interacting with an interactive molecular dynamics (IMD) simulation, where three Buckminster fullerene ($C_{60}$) molecules were simulated in the same virtual space, all with different bond force constants. The force constants affected how rigid the molecule appeared in VR: a large force constant meant the molecule was stiff and inelastic, and a small force constant meant the molecule was more flexible. Most participants were able to notice a difference between how the molecules behaved in the simulation, and correctly rank the $C_{60}$ molecules according to 'elasticity' (bond force constants). There is potential for the sensation of pseudo-haptic feedback to be exploited for drug design. For example, an IMD-VR user could estimate the barrier to dissociation of a drug from a protein based on how 'easy' or 'difficult' it feels to remove the drug from the binding site.

# 4. Applications of VR relevant to drug discovery

Although this review focuses on interactive VR, non-interactive VR of biomolecular structures clearly has uses in drug discovery, as noted above. VR provides depth perception which improves how the molecules are perceived in 3D and, as the screens are directly in front of your eyes, VR offers advantage of being able to view molecules with an increased field of view (compared to viewing on a screen or through stereoglasses) and as simply as moving your head, making the technology excellent as a graphical display.

The developers of ProteinVR (an example of a protein viewed in ProteinVR is shown in Fig. 2.4) tested how visualising a protein-ligand complex in ProteinVR compares to a 2D molecular viewer (VMD) [75]. The complex chosen (*T. brucei* RNA editing ligase 1, known as REL1, and V2, a naphthalene-based inhibitor) has a binding pocket that lies in a deep, narrow cavity. The structure of the complex used had been generated using the automated docking program AutoDock Vina [76]. The users found that, by viewing the molecule in VR, it allowed intuitive exploration of the narrow binding pocket cavity by the user moving their head. They even hypothesized that the AutoDock Vina pose may be incorrect and suggested interactions which could be more favourable to binding. This is an excellent demonstration of how users can utilise their spatial awareness in VR to note irregularities in the structure that might otherwise go unnoticed on a 2D screen.

Molecules are fundamentally dynamic [77], and therefore it can be important to consider molecular motion when designing drugs, *e.g.,* to see how a protein and drug candidate interact with each other over time. MD simulations [78-81] are increasingly widely used in drug design and development. Popular programs currently used for visualizing molecular dynamics trajectories include VMD [75], Pymol [82], and Chimera [83]. However, as these are typically projecting 3D motion onto a 2D screen, information is lost. Interactive VR software such as Nanome [30] and ChimeraX [84,85] can visualize molecular dynamics trajectories (Fig. 2.1). Features such as the ability to play, pause, rewind, fast-forward, slow down, and speed up the simulation that are available in traditional molecular viewers are also available in VR software.

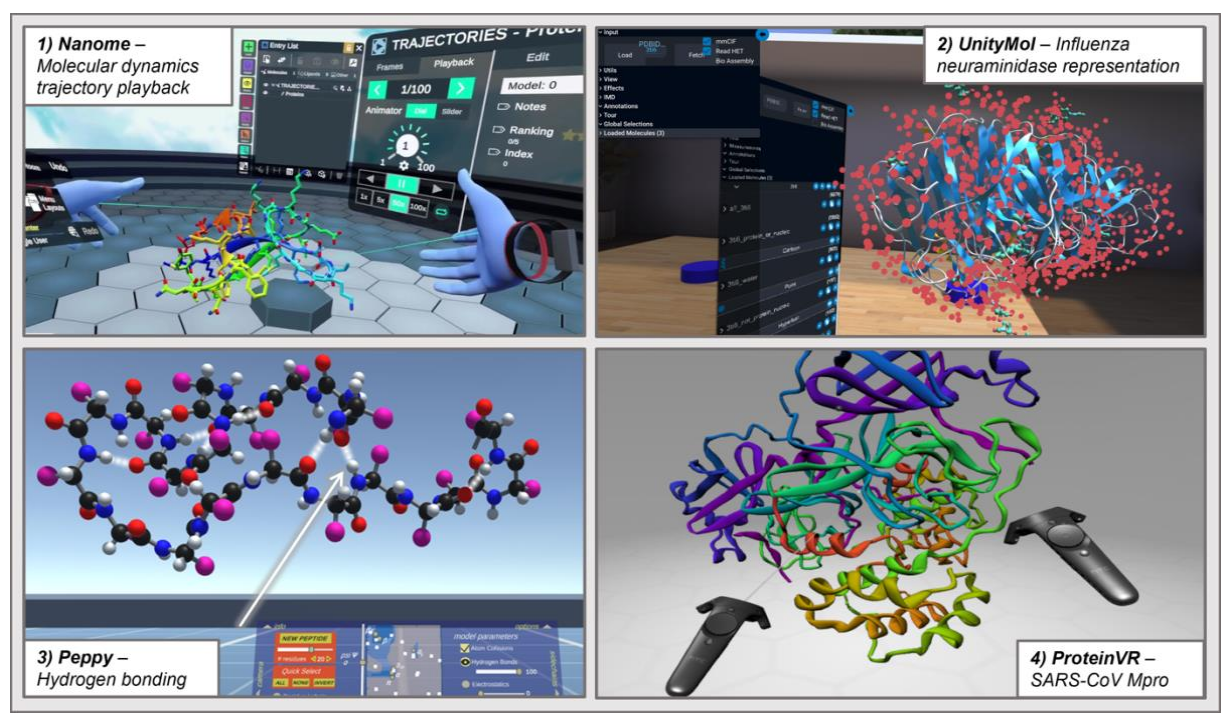

*Figure 2: **Scenes from VR software for molecular modelling.** 1) A scene from the interactive molecular dynamics tutorial from the Nanome software, where the user is inspecting the first frame of a trajectory. 2) UnityMol software showing a cartoon and hyperball representation of the influenza neuraminidase protein (PDB 3TI6). 3) Hydrogen bonding (rendered as white glowing interactions) between a 20 amino acid long polypeptide in the software, Peppy. 4) The SARS-CoV Mpro (PDB 2Q6G) in a rainbow cartoon representation using the web-based VR software ProteinVR.*

Just as with more standard screen-based GUIs, it is useful to visualise molecular properties and interactions such as hydrogen bonds and the electrostatic potential of the molecule(s) in VR. Electrostatic interactions such as hydrogen bonds, salt bridges, and charge-dipole interactions contribute greatly to protein structure and stability, and ligand binding [86,87]. Drug binding relies on favourable interactions between the drug and the receptor, therefore viewing and understanding these interactions is crucial to drug design and development [88,89]. Many interactive VR programs such as Molecular Rift, Nanome, and Peppy include a toggle for switching hydrogen bond visualisation on and off. By viewing these interactions in 3D, the user can utilise their depth perception to inspect the distance and angle between interacting groups. This is valuable for suggesting where modifications could be made to the drug or protein to generate additional hydrogen bonds. Molecular Rift and Peppy offer a striking visualisation of this type of interaction, portraying the contact between the donor and acceptor atoms as a bright, dynamic cloud (Fig. 2.3). In contrast to the flat dotted line, which is often used to indicate a hydrogen bond in 2D molecular visualisation software, this makes it obvious to the user when a hydrogen bond is present because it is visually distinct from the rest of the biomolecule.

Nanome (version 1.22.1) also provides the option to measure the distance and angle manually in VR, by clicking on the measuring tool and physically selecting atoms of interest. The value of the distance or angle is then displayed above the selected atoms. Finally, the ability to simply move your head and inspect the molecule from a different angle may be helpful when looking at hydrogen bonds (or other non-covalent interactions) between atoms buried below the surface of the molecule, where the structure is more labyrinthine.

In addition to visualising electrostatic interactions, it may also be important view the overall electrostatic potential of the molecule(s), as this can aid predictions in how and where a drug may bind to a receptor [90]. The visualisation of electrostatic potentials using computer graphics was first introduced by Weiner *et al.* in 1982 [91], who proposed that this kind of visualisation may be beneficial for drug design. Indeed, the visualisation of electrostatic surfaces has proven to be useful for investigating protein-ligand association for CADD [92-94]. Software such as adaptive Poisson-Boltzmann solver (APBS) [95,96] and PDB2PQR [97] facilitate electrostatic calculations of biomolecules for visualisation.

The developers behind UnityMol (example of interface shown in Fig. 2.2) have created a VR platform for simultaneous calculation and visualisation of electrostatic properties of molecules, named UnityMol-APBS [98]. The advantages that this programme has over APBS with a 2D viewer is that the user (i) has increased depth perception and field of view for examining the molecule, and (ii) can prepare APBS input files from within the virtual interface, using a graphical user interface (GUI) to send the input files directly to APBS tools. By simply reaching out, selecting the atoms of interest, and then clicking a button in the GUI, the user can seamlessly perform calculations within VR using the interface with APBS. The combination of the virtual interface with the APBS toolkit significantly reduces the time needed to manually prepare input files, which can often be verbose and rely on the user having prior knowledge of the software's language and formatting style. Here, the need for learning a completely new software is removed and replaced by a few easy gestures.

To investigate the utility of VR for exploring electrostatic surface potentials, the UnityMol-APBS developers examined a particular enzyme (*Torpedo californica* acetylcholinesterase, known to operate quickly due to its electrostatic properties) and a substrate molecule. In VR, with a large field of view and ability to translate, rotate, and scale the size of the molecule, it was immediately clear from the cluster of animated electrostatic field lines—which represent the electrostatic potential gradient—where the active site was situated. The immersive nature of VR also provides a sense of spatial awareness, meaning the user can grab the substrate and translate it as a rigid body through the binding cavity. In this way, users can get a sense of how feasible the pathway is.

## 5. On-the-fly modification

Drug design has been described as both a science and an art [99], and therefore, creativity and ingenuity are integral to success. A study investigating the relationship between creativity and factors relating to creativity (such as flow of work and attention) of individuals tasked with a clothing design challenge found (via EEG measurements of brainwaves) that a VR environment allows participants to focus more on the challenge at hand. Users reported that the ability to walk about and observe their creation from different angles inspired new ideas for improvement, compared to designing their product with a pencil and paper [100,101]. Furthermore, the ability to interact with complex 3D structures in VR inspires creative discussion around what interactions contribute to drug binding, and what could be altered to improve these interactions [30]. Many programs used in CADD are focused on building and structural modification of small drug candidates and proteins. As the orientation of functional groups can be vital in terms of drugs binding, performing complex 3D modelling on a flat screen could lead to some error. This raises the question: what if we could perform these modifications 'on-the-fly' in a 3D space, obtaining immediate answers to our hypotheses and thus driving intellectual discussion?

Since the function of many proteins relies on the structure they adopt, the capacity to perform structural modifications could be enormously powerful [102]. The interactive VR chemistry educational tool Peppy allows such modifications and a cohort of undergraduate chemists used it to perform a range of 3D modelling tasks: from creating short chain polypeptides to constructing α-helices and β-sheets with complex hydrogen bonding networks to teach them the fundamentals of protein structure [103]. The study showed that students using interactive VR were enthusiastic and creative when building and scrutinizing polypeptides. Some students also reported that interacting with polypeptides in VR cleared up previous misconceptions they had from learning protein structure in 2D [103]. The developers of Peppy state that, while their primary aim is teaching secondary structure of proteins to undergraduate students, they also recognise its potential use in research [103]. They also highlight that the goal of designing a tool capable of on-the-fly modifications and mutations was to encourage engagement and creativity, as well as inspiring a deeper exploration of biomolecular structures than a 2D builder would allow. A limitation of this software is that it does not currently support larger proteins (currently, the maximum is 32 residues). Nanome allows modifications of much bigger proteins.

The ability to make on-the-fly modifications to both the internal structure (such as rotating functional groups) and the chemical structure (like mutating a functional group) allows rapid testing of hypotheses and inspires creative thinking, as previous studies of GUIs etc have shown [100,101]. Nanome [30] allows building and modifying drug-like molecules within a VR environment. Researchers were able to build a small molecule inhibitor from scratch in the active site of an RIP2 kinase (PDB 5W5O), achieving an RMSD of 1.8 Å with respect to the crystal structure [104] (around 2 Å is considered by docking programmes to be a sensible cut-off for finding the 'correct' pose). Furthermore, starting from the crystal structure, users were tasked with making chemical alterations to the drug, where the objective was to form additional hydrogen bonds with the enzyme to achieve a more tightly bound pose [30]. This shows that, not only were users of Nanome able to produce sensible drug-protein structures similar to crystal structures, but they could also make predictions for more favourable interactions and trial them directly. Nanome also provides an energy minimisation feature that can remove human induced artefacts, such as unrealistically long chemical bonds, and return a more energetically reasonable structure. In another test case, a VR session was held where four Nanome users (chemists) were tasked with pointing out sites for macrocyclization of an active starting compound [30]. The aim was to preserve activity of the compound while improving physicochemical properties. Within minutes the users had become accustomed to the software and were discussing how they could alter the structure they were given. The group were successful in creating structures with improved physicochemical properties, whilst retaining activity.

The interactive VR tool Narupa Builder (Fig. 3), an application of the interactive VR framework Narupa, developed at the University of Bristol, also allows the creation and manipulation of molecules. Fig. 3 shows a drug structure (oseltamivir carboxylate, an influenza neuraminidase inhibitor) and the steps taken by a user to create the drug in an interactive VR environment. While this small drug was generated atom by atom, Narupa Builder also has a built-in library that contains common molecular fragments, such as amino acids and aromatic rings, that allow the rapid building of much larger molecules. There is also an energy minimisation feature, like Nanome, and the ability to export the resulting structure to a MOL2 format. The Narupa Builder is freely available for download via https://irl.itch.io/narupa-builder.

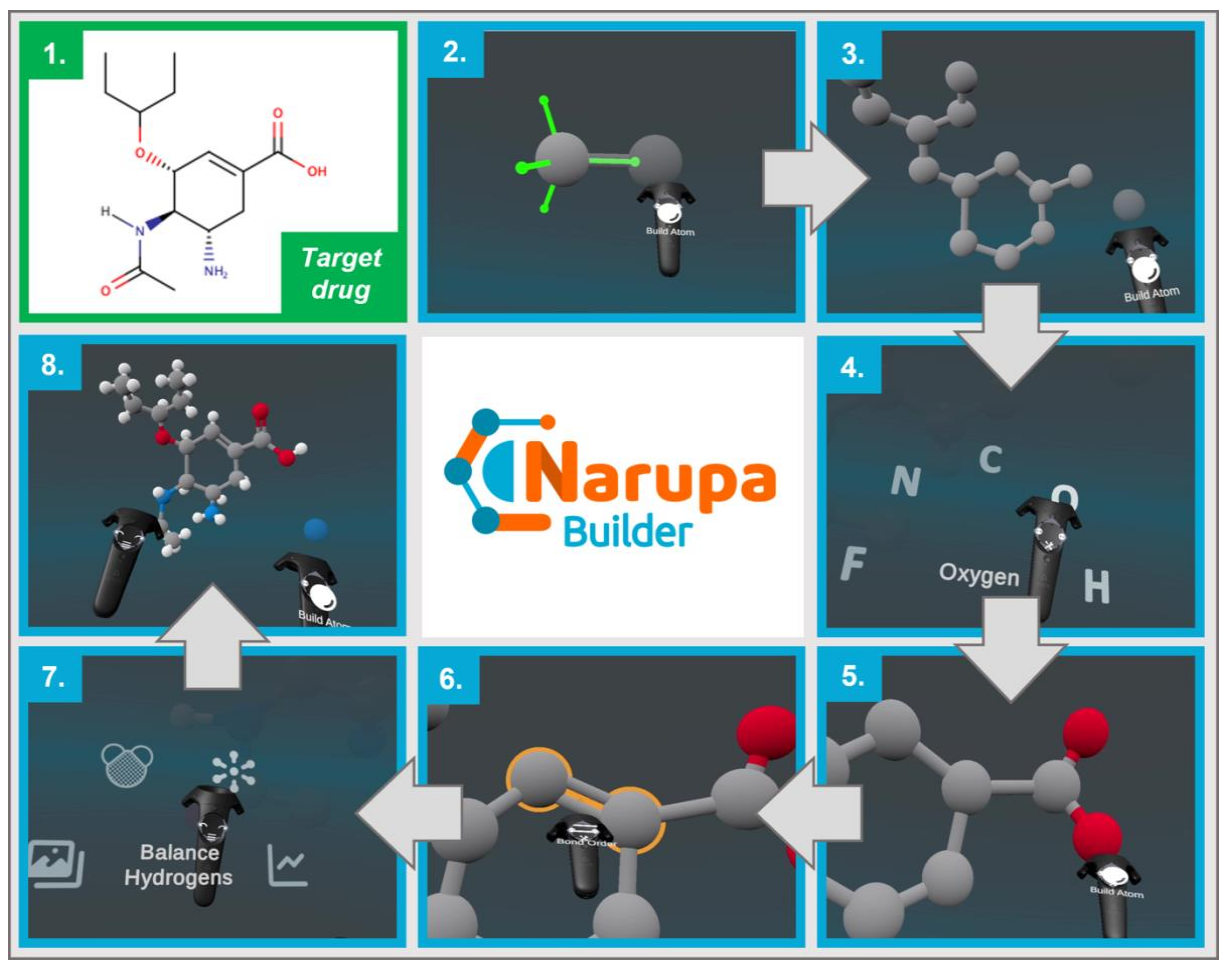

*Figure 3: **Constructing oseltamivir using the Narupa Builder**. 1) The structure of the target drug (oseltamivir carboxylate, commonly known as Tamiflu, an influenza drug) to be built using Narupa Builder. 2) Building the first carbon atoms of the 6 membered ring. The green axes represent the possible sites for a new atom to be bonded to the carbon. 3) The carbon backbone of the drug, without any hydrogens. 4) Changing from a carbon atom to an oxygen atom to be added to the drug. 5) Creating the carboxylate functional group. 6) Changing the bond order between two carbon atoms from a single to a double bond. 7) Finally, adding hydrogens to all the possible sites of hydrogenation. 8) The resulting structure.*

# 6. Interactive Molecular Dynamics in Virtual Reality (IMD-VR)

One of the most exciting recent developments in this area is the combination of interactive molecular dynamics simulations with virtual reality (IMD-VR) [31], allowing researchers to literally reach into their MD simulations and orchestrate the pathway of molecular motion [105]. IMD [106-111] has some similarities to steered MD (SMD) [112] which uses small forces applied over a long simulation time to simulate atomic force microscopy (AFM) experiments or biological structural change. IMD differs as large forces are applied by the user, and the experiments are very fast to run at the cost of a decrease in accuracy [113]. In IMD, the user steers the system to drive rare/high barrier events such as protein conformational changes and ligand binding. Various scientific problems have been tackled with IMD including: ligand binding [114], protein engineering [115], polymers [107], colloids [108], transport through membrane proteins [116], fitting X-ray crystallography [111], education [117,118] and even small ~(40 atoms) *ab initio* calculations [119]. However, performing complex 3D tasks with a 2D interface can be challenging. Although IMD had been around during the early VR era [107], the breakthrough year was 2001 when Stone, Gullingsrud, and Schulten [106] interfaced the graphical display capabilities of VMD with the molecular mechanics/dynamics engine NAMD [120] and contemporary others did IMD in java [108]. Interactivity was recognised as important, and since interaction with the simulation is via applied forces, further work in this area concentrated on adding haptic interfaces [106,113,115]: for example, force-feedback gloves combined with MD simulations via IMD were used in VR to design proteins for their stiffness and mechanical properties [115]. Some groups even combined IMD, VR, and haptics [114,116], however, those software permit only one user at a time to interact with the simulation.

The IMD-VR software Narupa IMD [32] allows users to interact with MD simulations (utilising the OpenMM physics engine [121]) on-the-fly in a fully immersive VR space. The purpose of OpenMM here is to propagate the MD; that is, the user supplies the force that moves the atoms, and OpenMM numerically integrates Newton's equations of motion and provides the new positions of the atoms. For every timestep, this process is repeated until we have a trajectory that an IMD-VR user has effectively produced. This means that the user can interact directly with molecules in VR, manually applying forces that propagate the simulation. It should be noted that other force engines can be used with Narupa IMD, users do not have to use OpenMM. In a recent human-computer interaction (HCI) study [53], participants were tasked with performing a range of molecular manipulation assignments (threading methane through a nanotube, changing helical screw sense, and tying a knot in a protein) using either a mouse and keyboard, a touchscreen, or IMD-VR. The participants were able to complete these complex 3D tasks faster using IMD-VR and with a higher success rate. IMD-VR has also been used as an undergraduate chemistry teaching tool, where students were tasked with performing small chemical rearrangements and docking tasks [122]. Feedback from students suggested that visualizing molecules in an interactive VR setting was more engaging than the traditional computer and mouse set-up they had previously used and allowed them to grasp chemistry concepts more quickly.

Automated docking of small molecules, fragments, and potential drugs is a popular CADD method due to its capacity to rapidly predict binding poses of small molecules to a target receptor molecule [123-126]. This provides insight into both the structural orientations (binding poses) of the small molecules as well as the corresponding binding affinities. Docking flexible molecules still poses a challenge for CADD. Deeks *et al.* investigated the use of Narupa IMD as an accurate approach for flexible, human driven, protein-ligand docking [127]. The experimental protocol involved testing whether expert and novice Narupa IMD users were able to interactively guide ligands (benzamidine, oseltamivir, and amprenavir) in and out of the binding pockets of three viral enzymes (trypsin, neuraminidase, and HIV-1 protease, respectively), and recreate their respective crystallographic protein-ligand binding poses (within 2.15 Å RMSD of crystal structure) in 5–10 minutes of real time. A bound structure from the interactive MD was then extracted and a short amount of MD was performed to test whether the structures created with IMD-VR were stable. By utilising human ability at 3D spatial manipulation, after only a short amount of time in VR (40 minutes, including time spent learning how to operate the VR equipment), non-expert users (most of whom were also not experts in drug docking) were able to produce docked structures of three viral enzymes with FDA approved drugs [127].

In a separate study [128], expert Narupa IMD users generated structures of a small molecule inhibitor and an 11-amino acid oligopeptide substrate complexed to the SARS-CoV-2 main protease (Mpro), an enzyme involved in replication of the virus that causes COVID-19 [129]. Various protocols were tested during the IMD-VR docking process, including running the simulation with and without backbone restraints on the enzyme. The expert users were able to produce structures that replicate experimentally observed crystal structures. The results emphasised the importance of formation of important hydrogen bonds and led to the recommendation to focus on forming such interactions to create stable bound structures. The expert users were also able to interactively dock the inhibitor and the substrate to an apo form of the Mpro, following the same procedure of forming key hydrogen bonds. The docked structures generated from apo Mpros remained stable in MD. A further study that compared docked structures of peptide-Mpro complexes generated from AutoDock CrankPep (a popular automated docking program),

Pymol, and IMD-VR showed that bound structures from all three methods were in good agreement with each other [130]. This was extremely encouraging.

Narupa IMD can save frames of the trajectory generated by the user. For example, if a user were to interactively grab a drug bound to an enzyme and pull it out into the solvent space, the user can save the frames of this trajectory using OpenMM, just like any other standard MD engine would allow. In a matter of minutes of real time, users can simulate a rare event such as drug unbinding [127,128,130]. Capturing these rare events may otherwise require a large amount of computational time and resources, as well as expert knowledge of the biological system prior to simulation. IMD-VR can allow users to sample a substantial range of the rugged energy landscape of a protein in an intuitive manner, using human spatial and chemical intuition to guide movements. IMD-VR allows manipulations on the atomic scale and observe how molecular systems evolve in real-time. This could provide an interesting first pass to how we expect a simulation to behave following a perturbation, *e.g.*, when a drug binds to a protein, how does the protein respond to allow the drug to bind? It should be noted that the forces that a user exerts on the simulation may be high (this is a non-equilibrium simulation). This should be considered when generating IMD-VR trajectories.

Currently, the biggest limitation of Narupa IMD is the number of atoms in the system that can be simulated and rendered at an appropriate framerate. While it is possible to reach several hundred thousand atoms for a *static structure* (where no MD engine is attached and we are just observing the structure, *e.g.*, LSD molecule bound to a membrane embedded HT2B human serotonin receptor using Narupa, see https://vimeo.com/420035802), dynamically rendering the positions of all these atoms as an MD simulation propagates is a current limitation. However, this also depends on the hardware being used for the simulation, and so advances in hardware will also improve issues with framerates.

IMD-VR has great potential to be used for drug design and development. As discussed above, the ability to observe how a molecule responds to a perturbation in real time could be a useful step in the drug discovery pipeline. For example, when manipulating a drug using IMD-VR, does a loop in the protein have to move for the drug to bind? Is there a key conformational changed that must happen? Perhaps a particular group of atoms in the drug can be modified to create more favourable interactions? Being able to simply reach out and create binding poses and pathways of drug binding on-the-fly, using human spatial and chemical intuition, has the potential to be a key step in drug development.

## 7. Conclusions

VR software will not replace other CADD methods; many of the features described in this review are already available in non-VR biomolecular manipulation software and have been useful for a long time. Instead, interactive VR offers an alternative way of modelling and manipulating biomolecules and can be used in conjunction with other CADD methods. For example, it may be useful to use VR to visualise a docked structure created from automated docking, employing the depth perception and immersion to deepen your understanding of the structure. Additionally, the interactive nature of the VR allows you to explore the structure in an intuitive manner, using simple gestures to grab and manipulate the system easily.

We encourage readers to try out these VR programs themselves. The ease of use of modern interactive VR programs, the immersion that they allow in 3D molecular space, and the ability to 'touch' molecules cannot be described adequately here; it's necessary to experience them. Despite the wealth of anecdotal evidence of VR tools improvement over non-VR visualisation and manipulation programs [131-133], there is a pressing need for well thought out and rigorously tested user studies and HCI experiments to quantify this improvement. In this review we have only mentioned VR; as headsets shrink in size, weight and cost, the related technology of augmented reality, AR, (VR with an optical pass-through such that the user sees both the virtual and real worlds simultaneously) will grow in popularity. It is easy to see how both collaborative VR and AR could be invaluable tools for teaching and research collaboration, with collaborative VR being especially useful for distance learning. The only modalities discussed at length here were vision, manual interactivity, and touch, there are more that can be added to VR. For example, InteraChem [117] is an interactive VR program built on Narupa IMD-VR for use as a teaching tool. It includes an interesting modality of atomic 'happiness' that encodes the energetic feasibility of a particular bonding arrangement and it is represented by emojis drawn on atoms. This approach co-opts the part of the brain that deals with human social interaction and makes high-level concepts incredibly easy to understand. Temperature feedback (of controllers or gloves) has also been raised as a possible method of information communication in IMD [115]. Colour is part of how we see the world and can be used to encode information easily and intuitively. VR also includes the ability to play sounds and here extra information can be passed to the user. There is room for significant creativity in including these aspects here: what colour is a wavefunction and how does this change as a molecule binds? How does the sound of a molecule's vibrations change as it goes through a transition state? Getting good answers to these questions will allow interactive VR to be even more intuitive for human users and the easier the technology is to use, the more we can harness human creativity efficiently.

# 8. Expert Opinion

Interactive VR is an emerging area that offers tremendous potential in drug design and development. State-of-the-art VR software and hardware can harness interactivity to manipulate and modify biomolecules on-the-fly, using human chemical and spatial intuition in practical and accessible ways. While humans cannot defeat computers in terms of speed and efficiency for producing results, human perception remains a vital part of drug design and discovery. The ability to be present within a molecular simulation and interact with it holds great promise.

Sophisticated VR hardware is now widely available and not very expensive. Currently, this includes HMDs and controllers. More 'natural' (*i.e.,* hand-like) controllers such VR gloves are emerging and allow a more tactile exploration of biomolecular structures. It is also possible to harness senses other than just sight, *e.g.,* touch: haptic feedback could be potentially very useful for drug design *e.g.,* feel the energy change and identify the bottleneck to pulling a drug out of a binding site. Other human senses such as sound could be utilised to aid exploring molecular systems, adding new ways to convey information about a model and its responses, and indeed this is being investigated [134-136].

A particularly interesting emerging frontier is IMD in VR. While IMD is by no means a new concept, the concept of humans interacting with real time MD simulations in a VR is still relatively new. Since molecules are inherently dynamic, molecular motion is important to consider when modelling biomolecules. Since some dynamics that are important to drug design occur on the order of nanoseconds to microseconds, *e.g.,* loop opening and drug binding, it may be useful to coax the system to explore areas of high energy using IMD-VR that may not otherwise be reached using standard MD simulations. While this kind of biased MD is not exclusive to VR (there are many non-VR enhanced sampling techniques available such as metadynamics, umbrella sampling, the string method, etc.), being able to physically reach out and manoeuvre molecules to find binding modes and pathways for drugs holds immense potential.

Finally, the collaborative nature of interactive VR means that many users can occupy the same virtual space, and even interact with the simulated molecules all at once. This means that scientists across the globe can inhabit the same simulation and observe the same molecular phenomena. This could be useful for sharing new drug candidates or protein structures with colleagues, where you can physically walk them through your findings, whilst also feeling like the other person(s) are in the same room as you.

Interactive VR is a rapidly developing area. The VR programs for molecular modelling mentioned in this review were published within the last 6 years, with the majority being published in the last 3 years. VR software and hardware will evolve rapidly. In the next ten years, it is likely that a VR setup will be a staple for drug designers, structural biologists, and medicinal and computational chemists. VR will also allow research teams to collaborate virtually. With the ease of viewing and manipulating 3D static and dynamical molecular properties, the ability to interact MD simulations, and the ability to change structures on the fly, interactive VR will be an increasingly important tool in the armoury of computational drug design and development.